\documentclass[prd,aps,floatfix,preprint,nofootinbib]{revtex4}
\usepackage[dvipdfmx]{graphicx,color}
\usepackage{epstopdf}
\usepackage{amssymb}
\usepackage{amsmath}
\usepackage{epstopdf}
\usepackage{subfigure}
\usepackage{epsfig}
\usepackage{float}

\def\simge{\mathrel{%
       \rlap{\raise 0.511ex \hbox{$>$}}{\lower 0.511ex \hbox{$\sim$}}}}
\def\simle{\mathrel{
       \rlap{\raise 0.511ex \hbox{$<$}}{\lower 0.511ex \hbox{$\sim$}}}}

\makeatletter
\newcommand{\figcaption}[1]{\def\@captype{figure}\caption{#1}}
\newcommand{\tblcaption}[1]{\def\@captype{table}\caption{#1}}

\newcommand{\no}{\nonumber}

\newcommand{\msbar}{\overline {\rm MS}}
\newcommand{\hlambda}{\hat{\lambda}}
\newcommand{\hg}{\hat{g}}
\newcommand{\hx}{\hat{x}}
\newcommand{\hy}{\hat{y}}
\newcommand{\hz}{\hat{z}}
\newcommand{\blambda}{\bar{\lambda}}
\newcommand{\brho}{\bar{\rho}}
\makeatother

\begin{document}

\title{
Linking $U(2)\times U(2)$ to $O(4)$ model via decoupling
}

\author{Tomomi Sato,\ Norikazu Yamada}
\affiliation{
        High Energy Accelerator Research Organization (KEK), %
        Tsukuba 305-0801, Japan\\
        Graduate University for Advanced Studies (SOKENDAI), %
        Tsukuba 305-0801, Japan}

\date{\today}

\begin{abstract}
 The nature of chiral phase transition of massless two flavor QCD
 depends on the fate of flavor singlet axial symmetry $U_A(1)$ at the
 critical temperature ($T_c$).
 Assuming that a finite $U_A(1)$ breaking remains at $T_c$, the
 corresponding three dimensional effective theory is composed of four
 massless and four massive scalar fields.
 We study the renormalization group flow of the effective theory in the
 $\epsilon$-expansion, using a mass dependent renormalization scheme,
 and determine the region of the attractive basin flowing into the
 $O(4)$ fixed point with a focus on its dependence on the size of the
 $U_A(1)$ breaking.
 The result is discussed from a perspective of the decoupling of massive
 fields.
 It is pointed out that, although the effective theory inside the
 attractive basin eventually reaches the $O(4)$ fixed point, the
 approaching rate, one of the universal exponents, is different from
 that of the standard $O(4)$ model.
 We present the reason for this peculiarity, and propose a novel
 possibility for chiral phase transition in two-flavor QCD.
\end{abstract}

\maketitle

\section{Introduction}
\label{sec:introduction}

Quantum chromodynamics (QCD) is a unique gauge theory in that its
nonperturbative phenomena are experimentally observable and thus one can
test our understanding on nonperturbative dynamics quantitatively.
Understanding the underlying principles of nonperturbative dynamics is
not only important in its own right but also interesting because it
could provide a solid basis for studying other hypothetical strong
coupling gauge theories.

In this paper, we address chiral phase transition of massless two-flavor
QCD at vanishing density.
This system is obviously different from QCD in real
world as it consists of massive flavors
\footnote{For the lattice studies of chiral transition of realistic 2+1
flavor QCD, see, for example,
Refs.~\cite{Bernard:2004je,Cheng:2006qk,Aoki:2006we,Bazavov:2011nk,Bhattacharya:2014ara,Jin:2014hea}.}, 
and hence studying this system may be considered to be academic.
On the other hand, since this system can be seen as one of extreme cases
of real QCD, precise knowledge on this system could provide with
foundations for understanding phase diagrams of real QCD as a function of
chemical potential, quark masses, or the number of flavors, etc.

The order of chiral phase transition of massless two flavor QCD has
been studied in uncountably many works both analytically and
numerically, but not settled yet~\cite{Vicari:2008jw}.
One of analytical methods is to examine the renormalization group (RG)
flow of the corresponding Landau-Ginzburg-Wilson (LGW) theory.
In 1983, Pisarski and Wilczek revisited the $\beta$ functions of linear
sigma models (LSMs) calculated in the $\epsilon$ expansion and
classified by the resulting RG flow the nature of chiral phase
transition of QCD with arbitrary number of massless
flavors~\cite{Pisarski:1983ms}.
However, the two flavor case remained uncertain because two distinct
effective theories are possible, depending on the presence of flavor
singlet axial [$U_A(1)$] symmetry at the critical temperature ($T_c$),
and they draw different conclusions.

In the case where a large $U_A(1)$ symmetry breaking remains at $T_c$,
$O(4)$ LSM should be analyzed.
$O(N)$ LSM has been well studied again both analytically and
numerically\footnote{See, for example,
Refs.~\cite{Kanaya:1994qe,Guida:1998bx,Antonenko:1998es,Berges:2000ew,Pelissetto:2000ek,Engels:2014bra}},
and the existence of the stable infrared fixed point (IRFP), or the
Wilson-Fisher fixed point, is established.

On the other hand, when the $U_A(1)$ symmetry is effectively and fully
restored at $T_c$, the symmetry of the system turns to
$U_L(2)\times U_R(2)$ (or $O(2)\times O(4)$).
This case has been also studied through various methods and is attracting
attention~\cite{Berges:1996ib,Berges:1996ja,Butti:2003nu,Delamotte:2003dw,Vicari:2007ma,Fukushima:2010ji,Pelissetto:2013hqa,Nakayama:2014sba,Grahl:2014fna}.
It appears that the nature of the transition in this system is still
under debate.

Numerical simulations based on lattice QCD can directly determine the
nature of the transition of massless two-flavor QCD without any
assumption, in principle.
Interestingly, a possibility of first order phase transition is recently
reported in one of the lattice calculations~\cite{Bonati:2014kpa} ,
while there remain many systematic uncertainties to be checked.

In this work, we will not pursue whether the $U_L(2)\times U_R(2)$ model
has an IRFP or not, and would rather focus on the case where the
$U_A(1)$ symmetry breaking is small but finite at $T_c$.
Although the size of the symmetry breaking at $T_c$ is determined by
nonperturbative dynamics and its precise value is not known yet, it is
probable from recent studies that the breaking effect is not
large~\cite{Bazavov:2012qja,Aoki:2012yj,Cossu:2013uua,Buchoff:2013nra,Bhattacharya:2014ara}.

This system is interesting from the field theoretical viewpoint.
$U_L(2)\times U_R(2)$ LSM contains eight degenerate scalar fields, and
by introducing the breaking, half of them gain mass proportional to the
breaking.
When the size of breaking is infinitely large, the system is simply
reduced to the $O(4)$ LSM and will end up with second order phase
transition~\cite{Pisarski:1983ms}.
Even if the breaking is tiny, we expect that the massive degrees of
freedom will decouple from the system and $O(4)$ LSM is eventually
realized as the flow goes into the infrared limit.
However, we are concerned that the decoupling
theorem~\cite{Symanzik:1973vg,Appelquist:1974tg} is not obvious in three
dimensions because the scalar quartic couplings have a mass dimension.

For example, four-point Green's functions can, in general,
have a term like $\bar{g}^2(P^2)/M^2$ due to massive fields with a mass
$M$, where $P$ represents a typical scale of external momenta and
$\bar{g}(P^2)$ is an effective quartic coupling connecting light and
heavy fields.
In three dimensions, $\bar{g}(P^2)$ has a mass dimension, and whether
$\bar{g}^2(P^2)/M^2$ vanishes in the $P^2\to 0$ limit is determined by
$P^2$ dependence of the running of $\bar{g}^2(P^2)$.
Indeed, the presence of non-decoupling effects is reported in
Ref.~\cite{Aoki:1997er}, where a theory with a dimensionful scalar
cubic coupling is examined in 3+1 dimensions.
It is thus interesting to see in the context of the RG flow how or even
whether the decoupling occurs.

We take the $\epsilon$ expansion approach to study this system since the
$\epsilon$ is suitable for investigating the detailed structure of the
decoupling on a fundamental level.
The calculation is done mainly in a mass-dependent renormalization
scheme such that $\beta$ functions contain information on finite mass of
would-be decoupling particles.
The consistency with the $\msbar$ scheme is checked through the
calculation of four-point correlation function.
As for other parts, we simply follow the standard.
With $\beta$ functions thus obtained, we determine the attractive basin
flowing into the $O(4)$ (or Wilson-Fisher) fixed point and see how the
area of the basin is affected by the size of the $U_A(1)$ breaking.

We point out that, although the effective theory starting from the
inside of the attractive basin eventually reaches the $O(4)$ fixed
point, one of the universal exponents turns out to differ from that of
the standard $O(4)$ LSM.
We present the reason for this peculiarity and propose a novel
possibility for chiral phase transition in two-flavor QCD, that is
second order phase transition with, say, the $U_A(1)$ broken scaling.

The same system has been studied in the functional renormalization group
(FRG) approach in Ref.~\cite{Grahl:2013pba}, where the phase transition
in the presence of a finite $U_A(1)$ breaking is concluded to be of first
order.
Since the $\beta$ functions calculated in the $\epsilon$ expansion are
embedded in the FRG, the same conclusion is naively expected to be
reached.
However, our conclusion is different from theirs.

Determining the order of the chiral phase transition of massless
two-flavor QCD has some impact on models of dynamical electroweak
symmetry breaking with electroweak baryogenesis.
For an attempt on the lattice, see
Ref.~\cite{Ejiri:2012rr,Ejiri:2014mada}.

Our analysis is performed at the leading order of the $\epsilon$
expansion.
Thus, our findings may be significantly affected by higher orders in the
expansion.
Furthermore, it is pointed out that the $\epsilon$ expansion is
sometimes not useful even for qualitative
discussions~\cite{Vicari:2007ma}.
Nevertheless, we believe that the $\epsilon$ expansion suffices for
exploring possible scenarios and making a survey of how the decoupling
of massive fields occurs along the flow toward the infrared limit.

The rest of paper is organized as follows.
In sec.~\ref{sec:model}, the effective theory we will discuss is
introduced.
We briefly summarize the leading $\epsilon$ expansion results for the
large and vanishing limits of the $U_A(1)$ breaking in
sec.~\ref{sec:e-expansion}.
The $\beta$ functions and the RG flow in the presence of a finite
$U_A(1)$ breaking are shown in sec.~\ref{sec:finite-cA}.
Based on those results, we determine the attractive basin in
sec.~\ref{sec:attractive_basin}.
The decoupling theorem is addressed in this system in
sec.~\ref{sec:decoupling}.
Summary and outlook are given in sec.~\ref{sec:summary}.
A part of this work has been published in Ref.~\cite{Sato:2013tka}.

\section{Effective theory}
\label{sec:model}

We take a linear sigma model (LSM) that has the same global symmetry as
that of massless two-flavor QCD around the critical temperature, $T_c$.
Following the standard procedure, we make a working hypothesis that the
system undergoes second order phase transition.
Then, the order parameters suitably chosen are small and hence is
used as an expansion parameter to construct Landau-Ginzburg-Wilson (LGW)
field theory.
At the critical temperature, the system becomes infrared conformal, and
modes with a divergent correlation length arise.
Then, the original system defined in four space-time dimensions can be
approximately described in three space dimensions.
In the following, the calculation is done in $D=4-\epsilon$ dimension,
and in the end $\epsilon=1$ is substituted.

The building block of the LSM is a $2 \times 2$ complex matrix field
\begin{align}
    \Phi
=   \sqrt{2}(\phi_0 - i \chi_0) t_0
  + \sqrt{2}(\chi_i + i \phi_i) t_i,
 \label{eq:Phi}
\end{align}
where $t_0=1_{2\times 2}/2$ and $t_i$=$\sigma_i/2$ ($i=1,2,3$) is the
generator of $SU(2)$ group.
$\phi_0$ and $\phi_i$ correspond to $\sigma$ and $\pi_i$ in more
commonly used name, respectively.
Similarly $\chi_0$ and $\chi_i$ to $\eta'$ and $\delta_i$.
Thus, $\chi_0$ denotes the iso-singlet pseudoscalar, and $\chi_i$ the
iso-triplet scalar.
Under chiral and $U_A(1)$ transformations, $\Phi$ transform as
\begin{eqnarray}
 \Phi \to e^{2i\theta_A}L^{\dagger}\Phi R\ \
 (L\in SU_L(2),\ R\in SU_R(2),\ \theta_A\in \rm{Re}).
 \label{eq:transformation}
\end{eqnarray}
$U_V(1)$ symmetry corresponding to the baryon number conservation was
omitted.
Since $\Phi$ can be considered as the order parameter of chiral
symmetry, nonzero vacuum expectation value of $\Phi$ indicates
spontaneous chiral symmetry breaking (S$\chi$SB).
Most general renormalizable Lagrangian conserving chiral and $U_A(1)$
rotations is then given by
\begin{eqnarray}
  \mathcal{L}_{U(2)\times U(2)}
= \frac{1}{2}\mathrm{tr}
  \left[\partial_{\mu}\Phi^{\dagger}\partial_{\mu}\Phi \right]
  +\frac{1}{2}m_0^2\,\mathrm{tr}\left[\Phi^{\dagger}\Phi \right]
  +\frac{\pi^2}{3}g_1\left(\mathrm{tr}[\Phi^{\dagger}\Phi]\right)^2
  +\frac{\pi^2}{3}g_2\mathrm{tr}\left[(\Phi^{\dagger}\Phi)^2\right].
\label{eq:U(N)}
\end{eqnarray}
which is referred to as $U(2)\times U(2)$ LSM.
Since we are interested in the system at around $T_c$, $m_0$ will be set
to zero in the analysis of $U(2)\times U(2)$ LSM.

In order to incorporate the effect of $U_A(1)$ symmetry breaking into
the system, the following terms are added
\begin{eqnarray}
      \mathcal{L}_{\rm breaking}
&=& - \frac{c_A}{4}(\mathrm{det}\,\Phi+\mathrm{det}\,\Phi^{\dagger})
    + \frac{\pi^2}{3}x\, \mathrm{Tr}[\Phi\Phi^{\dagger}]
      (\mathrm{det}\,\Phi+\mathrm{det}\,\Phi^{\dagger})
    + \frac{\pi^2}{3}y\,
      (\mathrm{det}\,\Phi+\mathrm{det}\,\Phi^{\dagger})^2\no\\
&&  + w \left(   \mathrm{tr}\left[\partial_{\mu}\Phi^\dagger\,t_2\,
                             \partial_{\mu}\Phi^*\,t_2 \right]
               + {\rm h.c.}
        \right).
\label{eq:anom}
\end{eqnarray}
The third term is symmetric under $Z_4$, and so is the rest under $Z_2$.
Rewriting the total Lagrangian in terms of the component fields, we
obtain
\begin{eqnarray}
    \mathcal{L}_{\rm total}
&=& \mathcal{L}_{U(2)\times U(2)}+\mathcal{L}_{\rm breaking}
\notag \\
&=&   (1+w)\frac{1}{2}(\partial_{\mu}\phi_a)^2
    + (1-w)\frac{1}{2}(\partial_{\mu}\chi_a)^2
    +\frac{m_{\phi}^2}{2}{\phi_a}^2+\frac{m_{\chi}^2}{2}{\chi_a}^2
\notag \\
& & + \frac{\pi^2}{3}
      \left[
         \lambda({\phi_a}^2)^2+(\lambda-2x)({\chi_a}^2)^2
       + 2(\lambda+g_2-z){\phi_a}^2{\chi_b}^2
       - 2g_2(\phi_a\chi_a)^2
      \right],
 \label{eq:Lfull}
\end{eqnarray}
where $\lambda=g_1+g_2/2+x+y$, $z=x+2y$ and $a$ runs 0 to 3.
We refer to the theory of eq.~(\ref{eq:Lfull}) as the $U_A(1)$ broken
LSM.
The non-zero value of $w$ affects all the terms through the redefinition
of the field normalization.
In the following, we set the tree level value of $w$ to zero, although
$w$ receives radiative corrections at two or higher loops unless
both $c_A$ and $x$ are zero.
Notice that the $c_A$ term in eq.~(\ref{eq:anom}) separates off the
degeneracy between $\phi_a$ and $\chi_a$ as
\begin{align}
 \label{eq:mass2}
 m_{\phi}^2=m_0^2-\frac{c_A}{2},\ \
 m_{\chi}^2=m_0^2+\frac{c_A}{2}.
\end{align}
In order to reproduce the properties of QCD vacuum, $c_A$ is taken to be
positive.
Otherwise the parity or iso-vector symmetry is broken.
As usual, $T=T_c$ corresponds to $m^2_\phi=0$, which means that
only $\chi$'s have a mass of $m_{\chi}^2=c_A>0$.
When $c_A$ is infinitely large, $\chi_a$ would be decoupled from the
system, and the total Lagrangian eq.~(\ref{eq:Lfull}) becomes $O(4)$
LSM,
\begin{align}
  \mathcal{L}_{O(4)}
=&\frac{1}{2}(\partial_{\mu}\phi_a)^2
 +\frac{\pi^2}{3}\lambda (\phi_a^2)^2.
 \label{eq:O(4)}
\end{align}

\section{RG flows for $c_A=0$ and $\infty$}
\label{sec:e-expansion}

In order to determine the renormalization group (RG) flow of the theory,
the $\beta$ functions in the effective theories are calculated.
Loop integrals are regularized by the dimensional regularization with
$D=4-\epsilon$.
In order to see the effects of the massive fields to the $\beta$
functions, we take a mass dependent renormalization scheme.
Here we choose the renormalization conditions that some specific
four-point amputated Green's functions should coincide, at a symmetric,
off-shell kinematic point (SYM) $s=t=u=\mu^2$, with their tree level
expressions:
\begin{align}
  \Gamma_4(\phi_1(p_1),\phi_1(p_2),\phi_2(p_3)\phi_2(p_4))|_{\rm SYM}
 &= -\frac{8}{3}\pi^2 \mu^\epsilon \hlambda_{\rm R}
 \label{eq:condition1}
\\
 \Gamma_4(\chi_1(p_1),\chi_1(p_2),\chi_2(p_3)\chi_2(p_4))|_{\rm SYM}
 &= -\frac{8}{3}\pi^2 \mu^\epsilon (\hlambda_{\rm R}-2\hx_{\rm R})
 \label{eq:condition2}
\\
 \Gamma_4(\phi_1(p_1),\chi_2(p_2),\phi_1(p_3)\chi_2(p_4))|_{\rm SYM}
 &= -\frac{8}{3}\pi^2 \mu^\epsilon 
     (\hlambda_{\rm R} + \hg_{2,{\rm R}} - \hz_{\rm R})
 \label{eq:condition3}
\\
 \Gamma_4(\phi_1(p_1),\chi_2(p_2),\phi_2(p_3)\chi_1(p_4))|_{\rm SYM}
 &= \frac{4}{3}\pi^2 \mu^\epsilon \hg_{2,{\rm R}}
 \label{eq:condition4}
\end{align}
where $p_{1,2}$ and $p_{3,4}$ are the incoming and outgoing momenta,
respectively.
$s=(p_1+p_2)^2=(p_3+p_4)^2$, $t=(p_1-p_3)^2=(p_2-p_4)^2$ and
$u=(p_1-p_4)^2=(p_2-p_3)^2$.
The conditions~(\ref{eq:condition1})-(\ref{eq:condition4}) are for the
$U_A(1)$ broken LSM.
Those for the $U(2)\times U(2)$ or the $O(4)$ LSM can be obtained by
simply omitting irrelevant couplings or conditions.
For example, the condition for the $O(4)$ LSM is given by
eq.~(\ref{eq:condition1}) only.
The mass dimension $\mu^\epsilon$ is factored out from the original
quartic couplings as explicitly shown, and the hatted couplings are
defined to be dimensionless.
Hereafter, the subscript ``R'' denoting renormalized one is omitted to
avoid notational complexity.

First we discuss the RG flow for the case with infinitely large $c_A$.
In this case, we deal with $O(4)$ LSM, eq.~(\ref{eq:O(4)}), which
contains only a single coupling $\hlambda$.
From the condition~(\ref{eq:condition1}), we obtain as the $\beta$
function~\cite{Brezin:1973jt}
\begin{eqnarray}
    \beta_{\hlambda,c_A=\infty}
 =  \mu\frac{d\hlambda}{d\mu}
 = -\epsilon\hlambda+2\hlambda^2.
 \label{eq:lambda-O(4)}
\end{eqnarray}
Although the $\beta$ function is known through higher orders in other
scheme~\cite{Antonenko:1998es}\footnote{See also
Ref.~\cite{Guida:1998bx}.},
we showed the one loop result for the later use.
$\hlambda$ reaches the IRFP $\hlambda_{{\rm IR},c_A=\infty}=\epsilon/2$
as long as the coupling at the initial scale $\Lambda$ satisfies
$\hlambda(\Lambda)>0$.
The existence of the IRFP meets the working hypothesis, and thus
massless two-flavor QCD satisfies the necessary condition for the second
order phase transition with the $O(4)$ scaling if $c_A$ is infinitely
large~\cite{Pisarski:1983ms}.

Next, we consider the case with $U_A(1)$ symmetry effectively restored.
$U(2)\times U(2)$ LSM in eq.~(\ref{eq:U(N)}) with $m_0=0$ contains two
independent couplings, $\hlambda=\hg_1+\hg_2/2$ and $\hg_2$.
With the conditions~(\ref{eq:condition1}) and (\ref{eq:condition4}),
their $\beta$ functions are obtained as~\cite{Pisarski:1983ms}
\begin{eqnarray}
      \beta_{\hlambda,c_A=0}
&=& - \epsilon\hlambda+\frac{8}{3}\hlambda^2
    + \hlambda \hg_2 +\frac{1}{2}\hg_2^2,
\label{eq:bl-cA0}
\\
      \beta_{\hg_2,c_A=0}
&=& - \epsilon \hg_2+2\hlambda \hg_2 + \frac{1}{3}\hg_2^2.
\label{eq:bg2-cA0}
\end{eqnarray}
The one loop $\beta$ functions~(\ref{eq:bl-cA0}) and (\ref{eq:bg2-cA0})
show no IRFP.
However, it should be noted that the existence of IRFP and hence
possibility of the continuous transition in $U(2)\times U(2)$ LSM is
reported in Refs.~\cite{Pelissetto:2013hqa,Nakayama:2014sba} employing
different approaches.

\section{RG flow for finite $c_A$}
\label{sec:finite-cA}

We now turn to the $U_A(1)$ broken theory~(\ref{eq:Lfull}) with a finite
and positive $c_A$.
The explicit one loop calculation yields
\begin{eqnarray}
     \beta_{\hlambda}
&=& - \epsilon\hlambda+2\hlambda^2
    + \frac{1}{6}f(\hat\mu)
      \left(   4 \hlambda^2 + 6 \hlambda \hg_2 + 3 \hg_2^2
             - 8 \hlambda \hz - 6 \hg_2 \hz + 4 \hz^2       
       \right),
\label{eq:bl}
\\
      \beta_{\hg_2}
&=& - \epsilon \hg_2+\frac{1}{3}\hlambda \hg_2
    + \frac{1}{3}f(\hat\mu) \hg_2
       \left( \hlambda - 2 \hx \right)
    + \frac{1}{3}h(\hat\mu) \hg_2
       \left( 4 \hlambda + \hg_2-4 \hz \right),
\label{eq:bg2}
\\
      \beta_{\hx}
&=& - \epsilon \hx+4f(\hat\mu)\left( \hlambda \hx -\hx^2 \right)
\notag \\
& & + \frac{1}{12} \left( 1-f(\hat\mu) \right)
      \left(  8 \hlambda^2 - 6 \hlambda \hg_2 - 3 \hg_2^2
            + 8 \hlambda \hz + 6 \hg_2 \hz - 4 \hz^2
      \right),
\label{eq:bx}
\\
      \beta_{\hz}
&=& - \epsilon \hz
    + \frac{1}{2}\left( 2 \hlambda^2 - \hlambda \hg_2 + 2 \hlambda \hz
                 \right)
    - \frac{1}{6}h(\hat\mu)
      \left(\,4\,\hlambda^2 + 3\,\hg_2^2 - 8\,\hlambda\,\hz + 4\,\hz^2
      \right)
\notag \\
& & + \frac{1}{6}f(\hat\mu)
      \left( -2 \hlambda^2 + 3\hlambda \hg_2 + 3 \hg_2^2
            - 2 \hlambda \hz - 6 \hg_2 \hz + 12 \hlambda \hx + 6 \hg_2x
            - 12 \hx \hz + 4 \hz^2\,
      \right),
\label{eq:bz}
\end{eqnarray}
where $\hat\mu=\mu/\sqrt{c_A}$ and
\begin{align}
    f(\hat\mu)
=   1 
  - \frac{4}{\hat\mu\sqrt{4+\hat\mu^2}}
    \arctan \sqrt{\frac{\hat\mu^2}{4+\hat\mu^2}},
\ \ \
    h(\hat\mu)
= 1 - \frac{1}{\hat\mu^2}\ln [1+\hat\mu^2]\, .
\end{align}
For small $\hat\mu$ these functions take the asymptotic forms,
\begin{eqnarray}
&& f(\hat\mu) = \frac{\hat\mu^2}{3} + O(\hat\mu^4),\ \
   h(\hat\mu) = \frac{\hat\mu^2}{2} + O(\hat\mu^4),
 \label{eq:zero_limit}
\end{eqnarray}
and for large $\hat\mu$,
\begin{eqnarray}
&& \lim_{\hat\mu\rightarrow \infty}f(\hat\mu)
 = \lim_{\hat\mu\rightarrow \infty}h(\hat\mu)
 = 1.
 \label{eq:infty_limit}
\end{eqnarray}
Thus, for infinitely large $c_A$ (or $\hat \mu\rightarrow 0$ with $\mu$ fixed),
$\beta_{\hlambda}$ [eq.~(\ref{eq:bl})] reduces to
$\beta_{\hlambda,c_A=\infty}$ [eq.~(\ref{eq:lambda-O(4)})] as expected.
On the other hand, in the $c_A\rightarrow 0$ limit (or
$\hat \mu\rightarrow \infty$ with $\mu$ fixed), the $\beta$ functions
eqs.(\ref{eq:bl})-(\ref{eq:bz}) agree with those in
Ref.~\cite{Aoki:2013zfa}, where the calculation is done with $c_A=0$ in
the mass independent scheme.
Note that the first term in each of eqs.(\ref{eq:bl})-(\ref{eq:bz})
comes from the mass dimension of the original dimensionful quartic
couplings.
Because of this, the dimensionless couplings behave like $1/\mu$ at the
tree level.

With the dimensional regularization, the wave function renormalizations
for $\phi$ and $\chi$ do not receive corrections at the one-loop.
We take the on-shell scheme in the renormalization of two-point
functions.
Thus, $\sqrt{c_A}$ is defined to be the pole mass of $\chi_a$ and does
not depend on the renormalization scale.

Two side remarks related to discrete symmetries are below.
Even if we set the mass of $\chi_a$ to zero ($c_A=0$) at tree level,
it would potentially receive radiative corrections unless $x$ is also
zero and $Z_2$ symmetry is present.
But the associated counter terms allow us to keep the renormalized
$c_A$ to zero.

Another remark is that $\hy=0$ at a certain scale can be kept at the
different scale only if $Z_2$ symmetry is preserved, {\it i.e.} both
$c_A$ and $\hx$ are zero.
We can explicitly check this in the $\beta$
functions~(\ref{eq:bl})-(\ref{eq:bz}).
These features are not affected by higher orders of the perturbation
series.

\begin{figure}[tbp]
 \begin{center}
  \begin{tabular}{cc}
  \includegraphics[width=0.5 \textwidth]{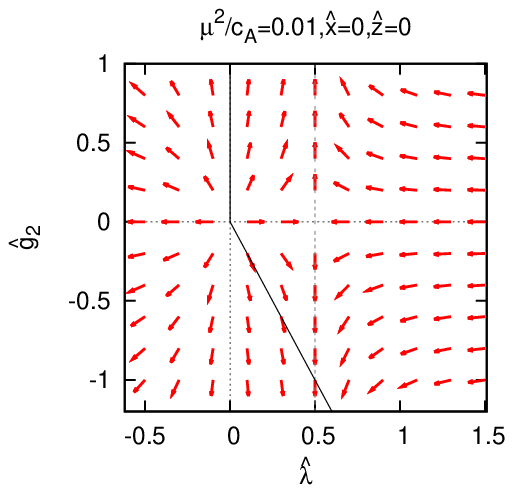} & 
  \hspace{-3ex}
  \includegraphics[width=0.5 \textwidth]{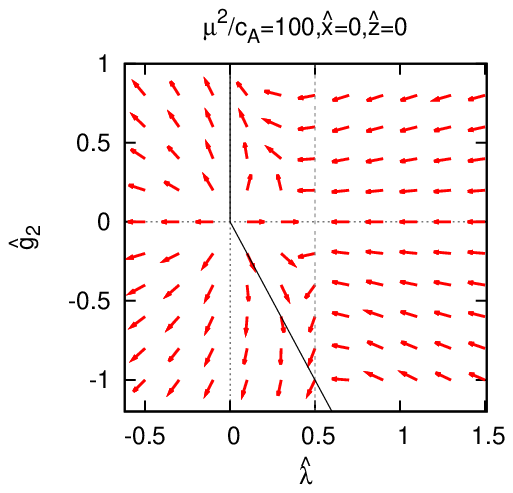} \\
  \end{tabular}
 \end{center}
 \caption{
 The RG flow of the couplings in the $U_A(1)$ broken
 LSM~(\ref{eq:Lfull}) projected on to the $\hlambda$-$\hg_2$ plane.
 $\mu^2/c_A$ is 0.01 (left) and 100 (right).
 The length of arrow does not represent the velocity of the flow.
 The solid lines show the stability bound obtained at the tree level
 analysis of the effective potential for the $U(2)\times U(2)$
 LSM~\cite{Ukawa:1995tc}.
 The dashed and dotted lines are just guide to eyes.
 }
 \label{fig:stream}
\end{figure}

The $\beta$ functions in (\ref{eq:bl})-(\ref{eq:bz}) indicate no stable
IRFP.
Fig.~\ref{fig:stream} shows an example of the RG flow in the $U_A(1)$
broken LSM with $\epsilon=1$, where the flow is projected on to the
$\hlambda$-$\hg_2$ plane for clarity.
In this example, $\hx$ and $\hz$ are set to zero everywhere.
The direction of the flow at each point is indicated by the arrow.
It turns out that at a region far from the line along $\hlambda=1/2$
the flow depends on $\mu^2/c_A$ only weakly while it is drastically
changed in the vicinity of the line for $\hg_2>0$.

\begin{figure}[tbp]
 \begin{center}
\begin{tabular}{cc}
   \includegraphics[width=0.5 \textwidth]{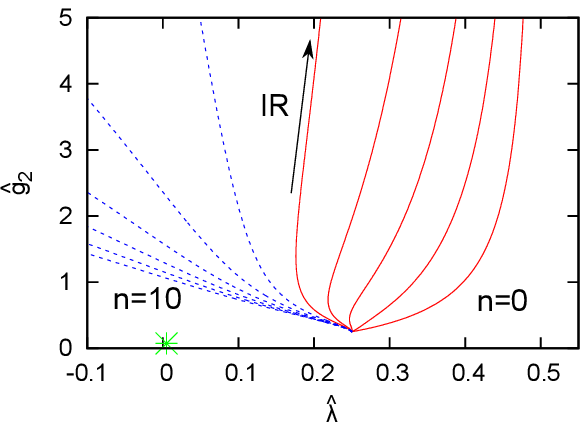}
 & \includegraphics[width=0.5 \textwidth]{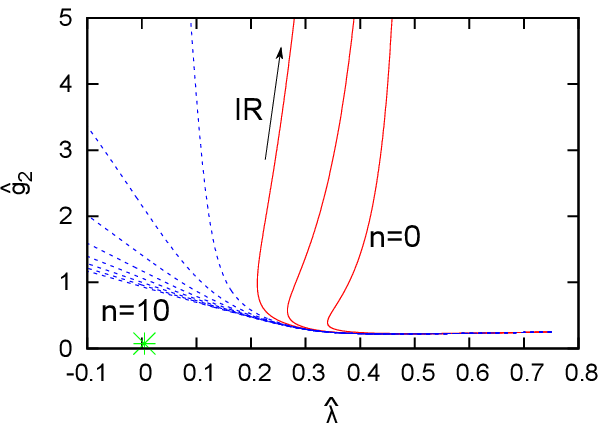} \\
\end{tabular}
 \end{center}
 \caption{
 The RG flow of the couplings in the $U_A(1)$ broken
 LSM~(\ref{eq:Lfull}) on the $\hlambda$-$\hg_2$ plane.
 Two initial conditions are chosen to be
 $(\hlambda(\Lambda),\ \hg_2(\Lambda),\ \hx(\Lambda),\ \hz(\Lambda))=
  (0.25,\ 0.25,\ 0,\ 0)$ and
 $c_A/\Lambda^2=\left(\frac{1}{2n+1}\right)^2$ (left), and
 $(0.75,\ 0.25,\ 0,\ 0)$ and 
 $c_A/\Lambda^2=\left(\frac{1}{10\,(2n+1)}\right)^2$ (right), as an
 example, where $n=0, \cdots, 10$.
 The IRFP of $U(2)\times U(2)$ LSM reported in
 Ref.~\cite{Pelissetto:2013hqa} is plotted at
 $(\hlambda,\ \hg_2)\sim$ (0.0048,0.073) (cross) as a reference.
 }
 \label{fig:flow}
\end{figure}

To see other aspects of the RG flow, the flow is calculated for two
initial conditions,
$(\hlambda(\Lambda),\ \hg_2(\Lambda),\ \hx(\Lambda),\ \hz(\Lambda))$
= $(0.25,\ 0.25,\ 0,\ 0)$ and $(0.75,\ 0.25,\ 0,\ 0)$
with varying $c_A/\Lambda^2$.
Fig.~\ref{fig:flow} shows the result projected onto the
$\hlambda$-$\hg_2$ plane, where the flows are classified into two types:
one approaching $\hlambda=1/2$ (solid curves) and the other going
$\hlambda=-\infty$ (dashed curves).
In the latter case (dashed curves), $\hg_2$ also diverges, {\i.e.} not
approaching some finite value, and then one usually expects first order
phase transition.

In the former case (solid curves), the flow never reaches an IRFP
because it does not exist, at least, at this order, but projecting it
onto the $\hlambda$-axis, it appears to reach the IRFP,
$\hlambda=\epsilon/2$.
In the infrared limit, $\mu^2/c_A$ becomes arbitrary small as long as
$c_A$ is finite.
Then $\chi$ would be effectively seen as a very massive field and
decoupled from the system.
Actually, $\hlambda=\epsilon/2$ is the IRFP of $O(4)$
LSM~(\ref{eq:O(4)}), which seems to support our interpretation that the
$U_A(1)$ broken theory~(\ref{eq:Lfull}) is reduced to the $O(4)$ LSM in
the IR limit via the decoupling of $\chi$.
This point is further discussed in the sec.~\ref{sec:decoupling}.

When approaching the $O(4)$ fixed point, $\hg_2(\mu)$ and $\hz(\mu)$
diverge as we will see below, but the terms including those couplings in
$\beta_{\hlambda}$ asymptotically vanish due to the suppression of
$f(\hat\mu)$ (see eq.~(\ref{eq:zero_limit})).
It means that although the couplings connecting $\phi$ and $\chi$
diverge the perturbative expansion of $\beta_{\hlambda}$ is still
sensible as long as this suppression works.

\begin{figure}[tbp]
 \begin{center}
  \begin{tabular}{cc}
  \includegraphics[width=0.5 \textwidth]{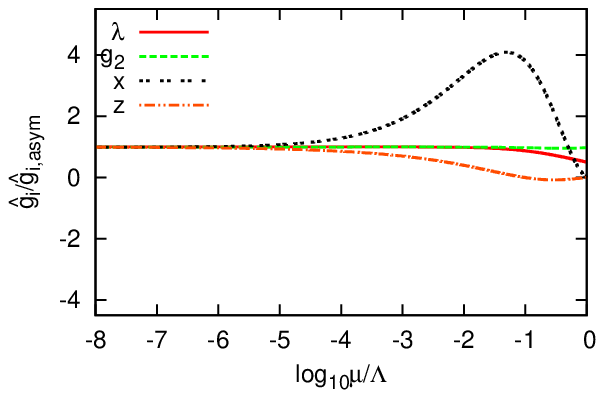}
& \includegraphics[width=0.5 \textwidth]{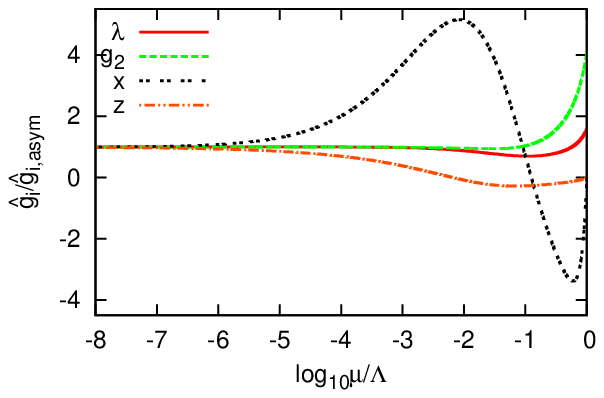} \\
  \end{tabular}
 \end{center}
 \caption{The $\mu$ dependence of the couplings is shown for two
 different initial conditions.
 Each coupling is normalized by its asymptotic behavior shown
 in eqs.~(\ref{eq:asymptotic_g2})-(\ref{eq:asymptotic_z}) and
 (\ref{eq:asymptotic_l}).
 The initial conditions are 
 $(\hlambda(\Lambda),\ \hg_2(\Lambda),\ \hx(\Lambda),\ \hz(\Lambda))=
 (0.25,\ 0.25,\ 0,\ 0)$ and $c_A/\Lambda^2=1$ (left), and 
 $(0.75,\ 0.25,\ 0,\ 0)$ and $c_A/\Lambda^2=0.01$ (right).
 The constant $c$ in eq.(\ref{eq:asymptotic_g2}) is $0.2613774$ and
 $0.4201792$, respectively.
 }
 \label{fig:asymptotic}
\end{figure}

It is interesting to note that the approaching rate to
$\hlambda=\epsilon/2$ differs from that in the ordinary $O(4)$ LSM
model.
In order to see this, we substitute $\lambda=\epsilon/2$ into
$\beta_{\hg_2}$, $\beta_{\hz}$ and $\beta_{\hx}$, and pick up the
dominant terms in the $\mu\rightarrow 0$ limit to obtain
\begin{eqnarray}
      \beta_{\hg_2}
&\approx&
    - \frac{5}{6}\,\epsilon\,\hg_2,
\label{eq:bg2-asymptic}
\\
      \beta_{\hx}
&\approx&
    - \epsilon {\hx}
    + \frac{1}{12}
      \left(\, - 3 \hg_2^2 + 6 \hg_2 \hz - 4 \hz^2 \,\right),
\\
      \beta_{\hz}
&\approx&
    - \frac{1}{2}\,\epsilon\,\hz - \frac{1}{4}\,\epsilon\,\hg_2,
\end{eqnarray}
where we have assumed that in the $\mu\rightarrow 0$ limit the terms
proportional to $f(\hat\mu)$ and $h(\hat\mu)$ are smaller than the other
terms.
Eq.~(\ref{eq:bg2-asymptic}) is easily solved, and the others too by
expressing the couplings as $\hz(\mu)\sim \mu^a$ and $\hx(\mu)\sim
\mu^b$ with unknown constants $a$ and $b$.
Then, the asymptotic behaviors of $\hg_2(\mu)$, $\hx(\mu)$ and
$\hz(\mu)$ in the vicinity of $\hlambda=\epsilon/2$ are found to be
related to each other as
\begin{eqnarray}
&& \hg_{2,{\rm asym}}(\mu)
=  \lim_{\mu\rightarrow 0} \hg_2(\mu)
=  c \left(\frac{\mu}{\sqrt{c_A}} \right)^{-5 \epsilon /6}
\label{eq:asymptotic_g2}
,\\
&& \hx_{\rm asym}(\mu)
 = \lim_{\mu\rightarrow 0}\hx(\mu)
 = \frac{3}{32}\hg_{2,{\rm asym}}^2(\mu), 
\label{eq:asymptotic_x}\\
&& \hz_{\rm asym}(\mu)
 =  \lim_{\mu\rightarrow 0}\hz(\mu)
 = \frac{3}{4}\hg_{2,{\rm asym}}(\mu),
\label{eq:asymptotic_z}
\end{eqnarray}
where the constant $c$ depends on the initial condition.
This behavior is consistent with the assumption above and confirmed in
the numerical calculation as shown in Fig.~\ref{fig:asymptotic}.

Substituting $\hlambda=1/2+\alpha$ and the asymptotic behavior
eqs.~(\ref{eq:asymptotic_g2})-(\ref{eq:asymptotic_z}) into
eq.~(\ref{eq:bl}), we obtain
\begin{eqnarray}
    \mu\frac{d\alpha}{d\mu} 
&\approx&
    \alpha
  + \frac{c^2}{24}\hat\mu^{2-\frac{5\epsilon}{3}}\,,
\end{eqnarray}
Then, as $\mu\rightarrow 0$, $\hlambda$ behaves like
\begin{eqnarray}
   \hlambda_{\rm asym}
=  \frac{\epsilon}{2} 
 - \frac{c^2}{8(5\epsilon -3)}{\hat\mu}^{2-\frac{5\epsilon}{3}}.
\label{eq:asymptotic_l}
\end{eqnarray}
The approaching rate in this case turns out to be $\sim \mu^{1/3}$ for
$\epsilon=1$ while in ordinary $O(4)$ LSM~(\ref{eq:O(4)}) it is linear
in $\mu$.
It is also interesting to note that $\hlambda$ always approaches
$1/2$ from below as demonstrated in Fig~ \ref{fig:asymptotic}.
This is not the case in the ordinary $O(4)$ LSM.
The origin of the discrepancy in the approaching rate is addressed in
sec.~\ref{sec:decoupling}.

\section{Attractive basin}
\label{sec:attractive_basin}

\begin{figure}[tb]
 \begin{center}
  \includegraphics[width=0.9 \textwidth,clip=true]{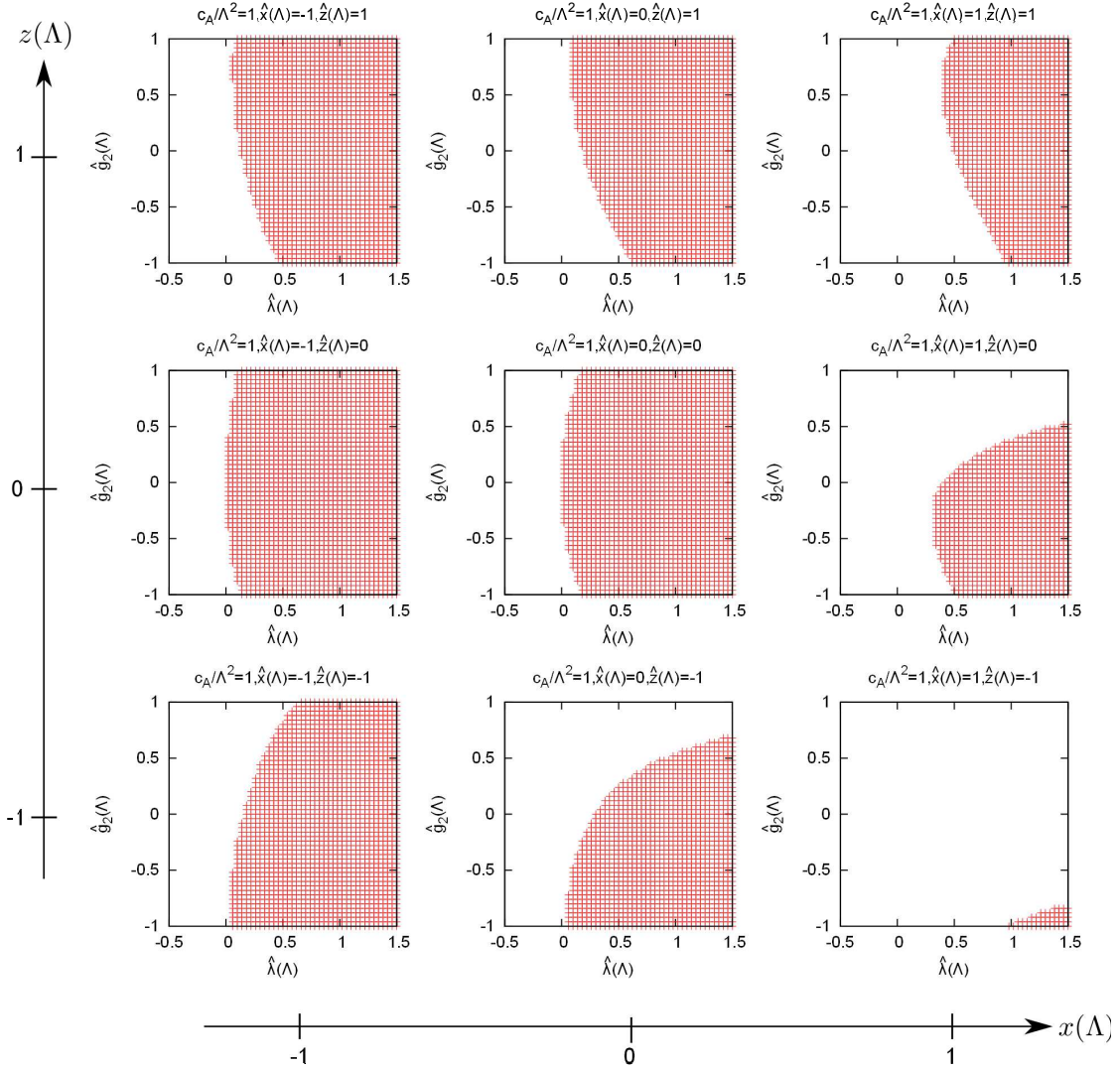}
 \end{center}
 \caption{The attractive basin in the
 ($\hlambda(\Lambda)$, $\hg_2(\Lambda)$) plane (hatched area) is shown,
 where $\hx(\Lambda)$ and $\hz(\Lambda)$ are varied from -1 to 1 as
 indicated.
 $c_A/\Lambda^2=1$.}
 \label{fig:reg1}
\end{figure}

\begin{figure}[tb]
 \begin{center}
  \includegraphics[width=0.9 \textwidth,clip=true]{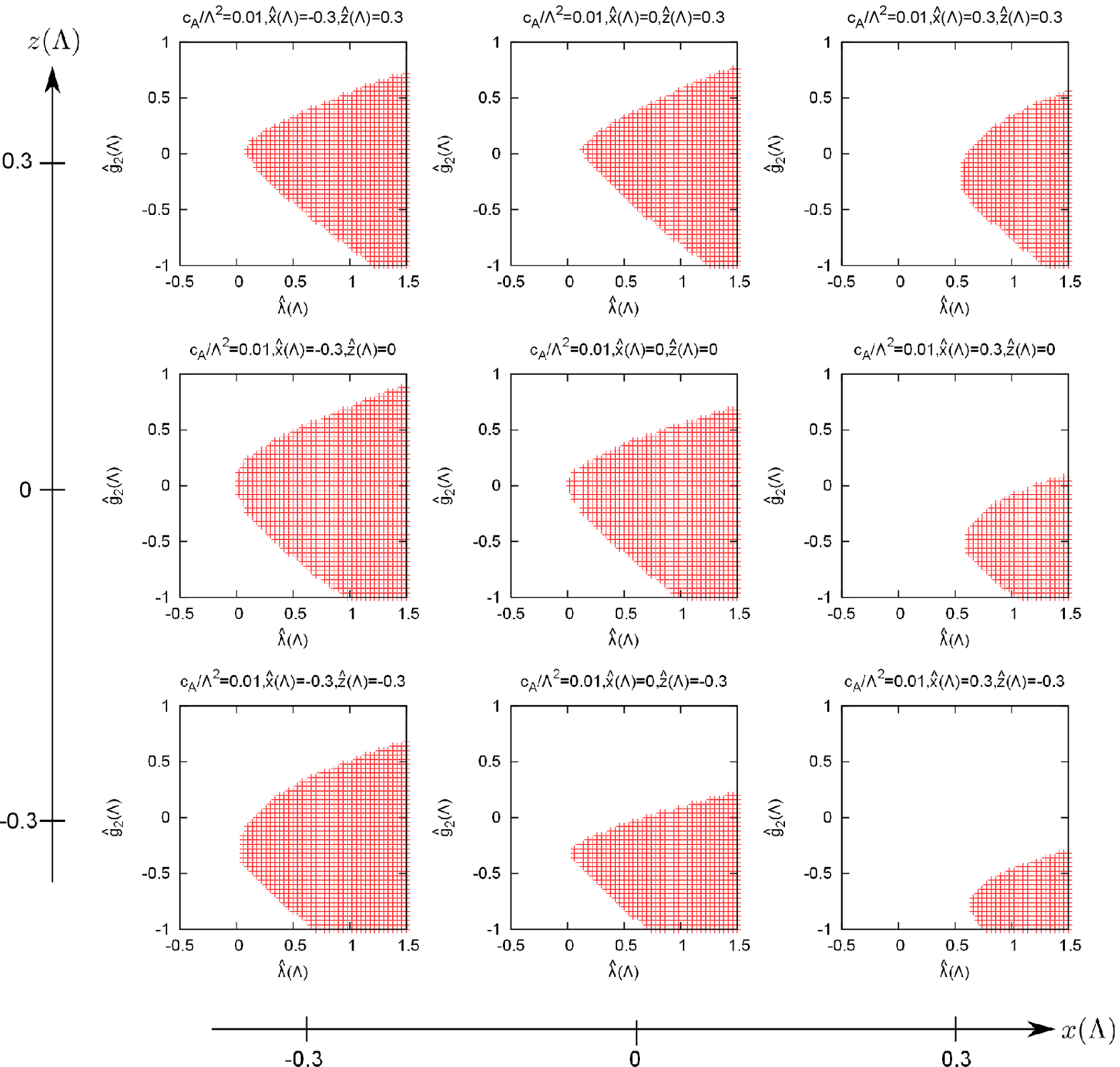}
 \end{center}
 \caption{The same plot as Fig.~\ref{fig:reg1} but for
 $c_A/\Lambda^2=0.01$.
 $\hx(\Lambda)$ and $\hz(\Lambda)$ are varied from -0.3 to 0.3.
 }
 \label{fig:reg2}
\end{figure}

Next, we present the attractive basin flowing into the $O(4)$ fixed
point.
We survey the initial coupling space on the
($\hlambda(\Lambda)$, $\hg_2(\Lambda)$) plane with two values of
$c_A/\Lambda^2=1$ and 0.01, shown in Figs.~\ref{fig:reg1} and
\ref{fig:reg2}, respectively.
The attractive basin is represented by the hatched area.
$\hx(\Lambda)$ and $\hz(\Lambda)$ are also varied as shown in the
figures.
It is seen that the attractive basin shrinks especially in the $\hg_2$
direction as $c_A/\Lambda^2$ decreases and is not very sensitive to
$\hx(\Lambda)$ and $\hz(\Lambda)$, unless $\hx(\Lambda)>0$ and
$\hz(\Lambda)<0$, in the region we studied.
Here let us assume that $\Lambda$ is the cutoff scale below which the
$U_A(1)$ broken LSM well describes massless two-flavor QCD and that the
size of $c_A$ is much smaller than $\Lambda$.
Then, in order for the $U_A(1)$ broken LSM to undergo second order phase
transition via the $O(4)$ fixed point, the initial condition, especially
$\hg_2(\Lambda)$, has to be suitably tuned.

\section{decoupling}
\label{sec:decoupling}

In this section, the decoupling
theorem~\cite{Symanzik:1973vg,Appelquist:1974tg} is revisited in this
system.
The theorem states that with a few
exceptions~\cite{Collins:1984xc,Aoki:1997er} the existence of heavy
particles is unknowable in low energy experiments as long as the
momentum scale is much smaller than the heavy particles' mass.
If the theorem holds in the present case, any $n$-point Green's
functions consisting only of $\phi_a$ in the $U_A(1)$ broken LSM should
agree with those in the ordinary $O(4)$ LSM in the infrared limit.
Thus, even if $\hlambda$ approaches the IRFP of the $O(4)$ LSM and the
$U_A(1)$ broken LSM appears to reduce to the $O(4)$ LSM, the observed
discrepancy in the approaching rate indicates that the decoupling
theorem does not hold in the $U_A(1)$ broken LSM.

To see this more explicitly, we calculate the four-point Green's
function of $\phi_a$ in the ordinary $O(4)$ and the $U_A(1)$ broken LSM.
In each LSM, the calculation is done with two renormalization schemes,
one being the symmetric scheme defined
in~(\ref{eq:condition1})-(\ref{eq:condition4}) and another being the
$\msbar$ scheme, to examine the scheme dependence.
The external momenta are set to $s=t=u=P^2$.
Since we consider the case where $P^2$ is extremely small, the RG
improvement is carried out.

\subsection{ordinary $O(4)$ LSM}
\label{subsec:O(4)}

First, we present the four-point function,
$G^{(4)}_{O(4)}(\{p_i\},\hlambda;\mu)$,
in the ordinary $O(4)$ LSM,~(\ref{eq:O(4)}).
Calculating it to one loop, and performing the RG improvement, which is
described in the next subsection in detail, one obtains
\begin{eqnarray}
    G^{(4)}_{O(4)}(\{p_i\},\hlambda;\mu)
&=& \left(\Pi_1^4 \frac{-1}{p_i^2}\right)^4
    P^{\epsilon} \mathcal{G}^{(4)}_{O(4)}(\blambda),
\end{eqnarray}
where
\begin{eqnarray}
      \mathcal{G}^{(4), {\rm sym}}_{O(4)}(\blambda)
= - \frac{8}{3}\pi^2\,\blambda,
&&\ \
      \mathcal{G}^{(4)\, \msbar}_{O(4)}(\bar{\lambda})
= - \frac{8}{3}\pi^2
      \left( \blambda - 2 \blambda^2 \right),
\end{eqnarray}
for symmetric and $\msbar$ scheme, respectively, and
$\blambda(P)$ satisfies
\begin{eqnarray}
    \frac{d\blambda(P)}{d\ln[P/\mu]}
= - \epsilon\blambda + 2 \blambda^2,
\end{eqnarray}
independently of the scheme at this order.
Then, the asymptotic behavior of the coupling in $P\to 0$ is given by
\begin{align}
    \bar{\lambda}(P\to 0)
\to \frac{\epsilon}{2}+c'\left(\frac{P}{\mu}\right)^{\epsilon}
\end{align}
with unknown constant $c'$, and hence those of the four-point function
\begin{eqnarray}
      \mathcal{G}^{(4)\, {\rm sym}}_{O(4)}(P\to 0)
&\to& - \frac{8}{3}\pi^2
      \left\{
       \frac{\epsilon}{2}+c'\left(\frac{P}{\mu}\right)^{\epsilon}
      \right\},\\
        \mathcal{G}^{(4)\, \msbar}_{O(4)}(P\to 0)
&\to& - \frac{8}{3}\pi^2
        \left\{
         \frac{\epsilon}{2}-\frac{\epsilon^2}{2}
          + c' \left(\frac{P}{\mu}\right)^{\epsilon}
        \right\},
\label{GO4_msbar}
\end{eqnarray}
are obtained\footnote{The $O(\epsilon^2)$ term in (\ref{GO4_msbar}) is
subject to the next to leading order.}.
Therefore, at the one loop, the approaching rate of the four-point
function of $\phi_a$ to its asymptotic value is $P^\epsilon$
and independent of renormalization scheme.

\subsection{$U_A(1)$ broken LSM with symmetric scheme}
\label{subsec:symmetric}

Next, we calculate the four-point function in the $U_A(1)$ broken LSM,
renormalized with the
conditions~(\ref{eq:condition1})-(\ref{eq:condition4}).
In the following, the couplings are, for convenience, rewritten as
\begin{align*}
 \lambda_1=\frac{\pi^2}{3}\lambda,\ \
 \lambda_2=\frac{\pi^2}{3}(\lambda-2x), \ \
 \lambda_3=\frac{2}{3}\pi^2(\lambda+g_2-z), \ \
 \lambda_4=-\frac{2}{3}\pi^2\,g_2,
\end{align*}
and $\rho=c_A/\mu^2$ is introduced.
To one loop, the four-point function is given by
\begin{eqnarray}
    G^{(4),{\rm sym.}}_1(\{p_i\},\{\hat{\lambda}_i\},\rho;\mu)
&=& \langle 0| \phi_1(p_1)\phi_1(p_2)\phi_2(p_3)\phi_2(p_4)|0\rangle
\no\\
&=& \left(\Pi_{i=1}^4\frac{-1}{p_i^2}\right) 
    \mu^{\epsilon} 
    g^{(4),{\rm sym.}}_1(P/\mu,\{\hat{\lambda}_i\},\rho),
\label{eq:defg^(4)_1}
\end{eqnarray}
where the dimensionless function
$g^{(4),{\rm sym.}}_1(P/\mu,\{\hat{\lambda}_i\},\rho)$ is
\begin{align}
     g^{(4),{\rm sym.}}_1(P/\mu,\{\hat{\lambda}_i\},\rho)
=& - 8\hlambda_1
   - \frac{1}{\pi^2}\int_0^1d\xi
     \biggl\{
      2^4{\hlambda_1}^2\ln[P^2/\mu^2]
      +2^2{\hlambda_1}^2(\ln[P^2/\mu^2]+\ln[P^2/\mu^2])
\notag\\ 
 &    +(\hlambda_3\hlambda_4+2{\hlambda_3}^2)
        \ln[\{\rho + \xi(1-\xi)P^2/\mu^2\}/\{\rho+\xi(1-\xi)\}]
\notag \\ 
 &    +2^{-2}{\hlambda_4}^2
        ( \ln[\{\rho+\xi(1-\xi)P^2/\mu^2\}/\{\rho+\xi(1-\xi)\}]
\notag \\ 
 &    +\ln[\{\rho+\xi(1-\xi)P^2/\mu^2\}/\{\rho+\xi(1-\xi)\}]
        )
     \biggr\}.
 \label{eq:G_1sym}
\end{align}
From the RG equation,
\begin{align}
\left[   \mu\frac{\partial}{\partial\mu}
       + \sum_i\beta_i\frac{\partial}{\partial\hat{\lambda}_i}
       + \beta_{\rho}\frac{\partial}{\partial\rho}
       + 4\gamma_\phi
 \right]\, G^{(4),{\rm sym.}}(\{p_i\},\{\hat{\lambda}_i\},\rho;\mu)
= 0,
\end{align}
that for $g^{(4),{\rm sym.}}_1(P/\mu,\{\hat{\lambda}_i\},\rho)$ is obtained as
\begin{align}
 \left[
   \frac{\partial}{\partial \ln[P/\mu]}
  -\sum_i\beta_i(\{\hat{\lambda}_i\},\rho) 
         \frac{\partial}{\partial\hat{\lambda}_i}
  -\beta_{\rho}(\{\hat{\lambda}_i\},\rho) \frac{\partial}{\partial\rho}
  -4\,\gamma_\phi(\{\hat{\lambda}_i\},\rho)-\epsilon
 \right]
  g^{(4),{\rm sym.}}_1(P/\mu,\{\hat{\lambda}_i\},\rho)
 =0,
\end{align}
where the derivative with regard to $\mu$ is altered to that of $P/\mu$.
Using the fact that $\gamma_\phi=0$ at the one loop, the solution is
given by
\begin{align}
 g^{(4),{\rm sym.}}_1(P/\mu,\{\hlambda_i\},\rho)
=& \mathcal{G}^{(4),{\rm sym.}}_1\left(\{\blambda_i(P) \},\brho(P)\right)
   \exp\left[
   \epsilon\, \int_0^{\ln[P/\mu]}d\ln[P'/\mu]
   \right]
\notag \\
=& \left(\frac{P}{\mu}\right)^{\epsilon}
   \mathcal{G}^{(4),{\rm sym.}}_1
     \left(\{\blambda_i(P)\},\brho(P)\right).
\label{eq:g^(4)_1solve}
\end{align}
Where $\blambda_i$ and $\brho$ satisfy
\begin{eqnarray}
&& \frac{d}{d\ln[P/\mu]}\blambda_i(P)
 = \beta_i(\{\blambda_i\},\brho), \ \
   \frac{d}{d\ln[P/\mu]}\brho(P)
 =-2\brho(P),
\end{eqnarray}
and the boundary conditions are set by
\begin{eqnarray}
&& \blambda_i(P=\mu)=\hlambda_i(\mu),\ \
   \brho(P=\mu)=\rho=c_A/\mu^2.
\end{eqnarray}
Then, we obtain, as the RG improved one,
\begin{eqnarray}
    \mathcal{G}^{(4),{\rm sym.}}_1(\blambda_i,\brho)
&=& - \frac{8}{3}\pi^2\,\blambda(P).
\end{eqnarray}

From the asymptotic behavior of $\bar{\lambda}(P\to 0)$, the asymptotic
behavior of the four-point function in $P\to 0$ is found to be
\begin{align}
      \mathcal{G}^{(4),{\rm sym.}}_1(\{\bar{\lambda}_i\},\bar{\rho})
\to - \frac{8}{3}\pi^2
    \left\{
     \frac{1}{2}-k\left(\frac{P}{\mu}\right)^{2-5\epsilon/3}
    \right\},
\end{align}
with a constant $k$.
Thus, in this scheme the asymptotic behavior of the four-point function
is that of $\blambda(P)$ as it should be.

\subsection{$U_A(1)$ broken LSM with $\msbar$ scheme}
\label{subsec:msbar}

To check the scheme dependence of the infrared behavior of the
four-point function, the calculation is repeated in $\msbar$ scheme.
$\beta$ functions in this scheme is easily obtained from
(\ref{eq:bl})-(\ref{eq:bz}) by putting $f(\hat\mu)=1$ and
$h(\hat\mu)=1$.
Thus, $\beta$ functions do not contain any information on the decoupling
by definition.
In this subsection, the couplings are defined in the $\msbar$ scheme
except for $\rho$, unless otherwise stated.
Following the same procedure in \ref{subsec:symmetric}, we obtain, as
the RG improved one,
\begin{align}
     \mathcal{G}_1^{(4)\,\msbar}(\{\blambda_i\},\brho)
=& - \frac{8}{3}\pi^2
     \left\{
      \blambda - 2 \blambda^2
     +\frac{1}{6} (
        4 \blambda^2 + 6 \blambda \bar{g}_2 + 3{\bar{g}_2}^2
      - 8\blambda \bar{z} - 6 \bar{g}_2 \bar{z} + 4\bar{z}^2
      )
    \right.
\notag \\ &
    \left.
      \times\frac{1}{2}\int_0^1 dx \ln[\brho+x(1-x)]
 \right\}.
 \label{eq:G4_msbar}
\end{align}
In contrast to the symmetric scheme, the $\chi$ mass ($\brho$)
dependence appears here.

Since we are interested in the $P$ dependence of
$\mathcal{G}_1^{(4)\,\msbar}$, we differentiate it with regard to
$\ln(P/\mu)$.
Neglecting higher order terms, it yields
\begin{eqnarray}
      \frac{d \mathcal{G}_1^{(4)\,\msbar}(\{\blambda_i\},\brho)}
           {d\ln[P/\mu]}  
&=& - \frac{8}{3}\pi^2
      \left\{
       \frac{d}{d\ln[P/\mu]}\blambda
     + \frac{1}{6}(
         4 \blambda^2 + 6 \blambda \bar{g}_2 + 3 {\bar{g}_2}^2
       - 8 \blambda \bar{z} - 6\bar{g}_2 \bar{z} + 4\bar{z}^2
      ) \right.
    \no\\
& & \left.
    \times\frac{1}{2}\frac{d \brho}{d\ln[P/\mu]}
          \frac{\partial}{\partial \brho}\int_0^1 dx \ln[\brho+x(1-x)]
    \right\}.
\end{eqnarray}
Now, using the followings,
\begin{eqnarray}
      \frac{\partial}{\partial \brho}\int_0^1 dx \ln[\brho+x(1-x)]
&=&   \frac{1}{\brho}\Bigl(1-f\left(1/\brho\right)\Bigr).
\\
      \frac{d}{d\ln[P/\mu]}\bar{\lambda}
&=& - \epsilon\bar{\lambda}
    + \frac{8}{3}\bar{\lambda}^2
    + \bar{\lambda}\bar{g}_2
    + \frac{1}{2}{\bar{g}_2}^2
    - \frac{4}{3}\bar{\lambda}\bar{z}
    - \bar{g}_2\bar{z}+\frac{2}{3}{\bar{z}}^2,
\end{eqnarray}
we obtain
\begin{eqnarray}
      \frac{d \mathcal{G}_1^{(4)\,\msbar}(\{\blambda_i\},\brho)}
           {d\ln[P/\mu]}
&=& - \frac{8}{3}\pi^2
      \left\{
      -\epsilon\bar{\lambda}+2\bar{\lambda}^2
      +\frac{1}{6}f\left(1/\bar{\rho}\right)
      (   4 \blambda^2 + 6 \blambda \bar{g}_2 + 3 \bar{g}_2^2
       -8\bar{\lambda}\bar{z}-6\bar{g}_2\bar{z}+4\bar{z}^2
      )
    \right\}
\no\\
&=&
  \frac{d \mathcal{G}_1^{(4)\,\mathrm{sym.}}(\{\blambda_i\},\brho)}
       {d\ln[P/\mu]}.
\end{eqnarray}
The last line holds because $\epsilon$ is counted as the same order as
the couplings.
Thus, it is confirmed that the $P$ dependence of the four-point function
agrees between two schemes.

\subsection{reason for the different approaching rate}

Here let us explore reasons for the different approaching rate.
The reason seems to be simply originating from the fact that the quartic
couplings describing interactions between the light $\phi_a$ and heavy
$\chi_b$ fields have a mass dimension in three dimensional theory.

The contribution of massive fields ($\chi_b$) with a mass $M$ to a
renormalized Green's function of light fields ($\phi$) at external
momentum $P$ will take the form of $\hg^2(P)P^2/M^2$ when 
$P^2/M^2\ll 1$, where $\hg$ represents a generic dimensionless quartic
coupling and is related to the coupling in Lagrangian as
$g=\mu^\epsilon\hg$.
This is indeed seen in eq.~(\ref{eq:G4_msbar}), if one expands the
logarithmic term assuming $1/\brho(P)=P^2/c_A \ll 1$.

If $D=4$ (or $\epsilon=0$), $\hg^2(P)P^2/M^2$ will vanish as $P^2\to 0$
because $\hg^2(P)$ depends on $P$, at most, logarithmically, but when
$D=3$ (or $\epsilon=1$), it does not in general because the factor $P^2$
can be compensated by $\hg^2(P)$, which behaves $\sim 1/P^2$ at the tree
level.
Thus, in general, the decoupling theorem does not hold when a coupling
has a mass dimension.
The same conclusion is reported in Ref.~\cite{Aoki:1997er}, where
non-decoupling effects of the scalar cubic interaction in 3+1 dimensions
is studied.

Another and more important reason is below.
Usually, the approaching rate is argued in terms of more familiar
quantity, $\omega$, defined by
\begin{eqnarray}
 \omega
= \frac{d\beta_{\hat\lambda}}{d\hat\lambda}
  |_{\hat\lambda=\hat\lambda_{\rm IRFP}},
\end{eqnarray}
which is one of the universal exponents.
The above results yield
\begin{eqnarray}
 \omega_{O(4)}=\epsilon\ \ \ \mbox{and}\ \ \
 \omega_{U_A(1) {\rm broken}}=2-5\epsilon/3,
\end{eqnarray}
for the $O(4)$ and the $U_A(1)$ broken LSM, respectively.

According to the general argument of renormalization group, $\omega$ is
determined by the RG dimension of the leading irrelevant operator in a
model under consideration.
While $({\phi_a}^2)^2$ is the one in the $O(4)$ LSM, it is not evident
in the $U_A(1)$ broken LSM but should not be the same as the $O(4)$ LSM
because $\omega_{O(4)}\ne \omega_{U_A(1) {\rm broken}}$.

One possible candidate is $(\phi_a\chi_a)^2$, which should become
eventually irrelevant since its effects to the low energy behavior is
expected to vanish as $\chi_a$ decouples from the system.
Since the coefficient of $(\phi_a\chi_a)^2$ term is $\hg_2$, we
calculate $\omega$ with $\hat{g}_2=0$ as a trial and obtain
$\omega_{\hg_2=0}=\omega_{O(4)}=\epsilon$.
Then, it is concluded from this observation that the operator
$(\phi_a\chi_a)^2$ effectively plays a role of the leading irrelevant
operator in the $U_A(1)$ broken LSM.
Therefore, the $U_A(1)$ broken LSM is the system which is invariant
under $O(4)$ rotation for $\phi_a$ in the IR limit, but does not obey
the $O(4)$ scaling.

It is important to notice that our study suggests a novel possibility
for the nature of chiral phase transition of two-flavor QCD.
Currently, three possibilities remains:
(i) first order
(ii) second order with the $O(4)$ scaling
(iii) second order with the $U(2)\times U(2)$ scaling.
We suggests the new one:
(iv) second order with, say, the $U_A(1)$ broken scaling.

\section{Summary and outlook}
\label{sec:summary}

The nature of the chiral phase transition of massless two-flavor QCD
depends on the fate of $U_A(1)$ symmetry at the critical temperature.
Two extreme cases with infinitely large and vanishing $U_A(1)$ breaking
have been well studied relying on effective theories and seem to have
their respective IRFP although the latter is not settled yet.
We have studied the case with a finite $U_A(1)$ breaking.

The RG flow of $U(2)\times U(2)$ LSM with a finite $U_A(1)$ breaking is
investigated in the $\epsilon$ expansion.
It turns out that if the couplings start from a certain region, {\it
i.e.} attractive basin, one of the couplings flows into the same fixed
point as the one in $O(4)$ LSM although the approaching rate is
different from the $O(4)$ case.
The interpretation of this is that the $U_A(1)$ broken LSM approaches
the $O(4)$ LSM in the IR limit via the decoupling of the massive fields.

The attractive basin flowing into the $O(4)$ fixed point shrinks as
$c_A$ decreases.
Thus, for smaller $c_A$, the phase transition of massless two flavor QCD
favors the first order phase transition more than the second.

The observed discrepancy in the approaching rate is caused by the
non-decoupling effect.
In other words, the decoupling rate of the massive fields is slower than
the approaching rate in the standard $O(4)$ LSM, and it effectively
changes the RG dimension of the leading irrelevant operator through
$(\phi_a \chi_a)^2$.
In order to establish the non-decoupling, it is clearly interesting to
calculate the other critical exponents and compare with those of the
$O(4)$ LSM.

The existence of an IRFP just satisfies a necessary condition for second
order phase transition.
The phase transition can be more clearly investigated by calculating the
effective potential.
Such a study is ongoing~\cite{Sato:2014effV}.

The analysis here consists of simple one-loop calculations, and hence the
results are neither quantitative nor conclusive.
Nevertheless, we believe that this simple analysis is still useful to
explore possible scenarios and offers a good starting point for further
study.

\section*{Acknowledgments}
We would like to thank G. Fleming, M. Hayakawa, Y. Nakayama, and
K. Kamikado for useful discussions and comments.


\begin{thebibliography}{99}

 \bibitem{Bernard:2004je} 
  C.~Bernard {\it et al.}  [MILC Collaboration],
  Phys.\ Rev.\ D {\bf 71}, 034504 (2005)
  [hep-lat/0405029].

\bibitem{Cheng:2006qk} 
  M.~Cheng, N.~H.~Christ, S.~Datta, J.~van der Heide, C.~Jung,
 F.~Karsch, O.~Kaczmarek and E.~Laermann {\it et al.},
  Phys.\ Rev.\ D {\bf 74}, 054507 (2006)
  [hep-lat/0608013].

 \bibitem{Aoki:2006we}
  Y.~Aoki, G.~Endrodi, Z.~Fodor, S.~D.~Katz and K.~K.~Szabo,
  Nature {\bf 443}, 675 (2006)
  [hep-lat/0611014].

\bibitem{Bazavov:2011nk} 
  A.~Bazavov, T.~Bhattacharya, M.~Cheng, C.~DeTar, H.~T.~Ding,
        S.~Gottlieb, R.~Gupta and P.~Hegde {\it et al.},
  Phys.\ Rev.\ D {\bf 85}, 054503 (2012)
  [arXiv:1111.1710 [hep-lat]].

\bibitem{Bhattacharya:2014ara} 
  T.~Bhattacharya, M.~I.~Buchoff, N.~H.~Christ, H.-T.~Ding, R.~Gupta,
        C.~Jung, F.~Karsch and Z.~Lin {\it et al.},
  Phys.\ Rev.\ Lett.\  {\bf 113}, 082001 (2014)
  [arXiv:1402.5175 [hep-lat]].

\bibitem{Jin:2014hea}
     X.~Y.~Jin, Y.~Kuramashi, Y.~Nakamura, S.~Takeda and A.~Ukawa,
     arXiv:1411.7461 [hep-lat].
         
 \bibitem{Vicari:2008jw}
 For previous efforts, see, for example,
  E.~Vicari and H.~Panagopoulos,
  Phys.\ Rept.\  {\bf 470}, 93 (2009)
  [arXiv:0803.1593 [hep-th]],
  and references therein.
         
\bibitem{Pisarski:1983ms}
  R.~D.~Pisarski and F.~Wilczek,
  Phys.\ Rev.\ D {\bf 29}, 338 (1984).

 \bibitem{Kanaya:1994qe}
  K.~Kanaya and S.~Kaya,
  Phys.\ Rev.\ D {\bf 51}, 2404 (1995)
  [hep-lat/9409001].

\bibitem{Guida:1998bx} 
 R.~Guida and J.~Zinn-Justin,
 J.\ Phys.\ A {\bf 31}, 8103 (1998)
 [cond-mat/9803240].

\bibitem{Antonenko:1998es}
 S.~A.~Antonenko and A.~I.~Sokolov,
 Phys.\ Rev.\ E {\bf 51} (1995) 1894
 [hep-th/9803264].
        
 \bibitem{Berges:2000ew} 
  J.~Berges, N.~Tetradis and C.~Wetterich,
  Phys.\ Rept.\  {\bf 363}, 223 (2002)
  [hep-ph/0005122].

\bibitem{Pelissetto:2000ek}
   A.~Pelissetto and E.~Vicari,
   Phys.\ Rept.\  {\bf 368}, 549 (2002)
   [cond-mat/0012164].

\bibitem{Engels:2014bra}
 J.~Engels and F.~Karsch,
 arXiv:1402.5302 [hep-lat].
         
\bibitem{Berges:1996ja} 
  J.~Berges and C.~Wetterich,
  Nucl.\ Phys.\ B {\bf 487}, 675 (1997)
  [hep-th/9609019].

\bibitem{Berges:1996ib} 
  J.~Berges, N.~Tetradis and C.~Wetterich,
  Phys.\ Lett.\ B {\bf 393}, 387 (1997)
  [hep-ph/9610354].

\bibitem{Butti:2003nu} 
  A.~Butti, A.~Pelissetto and E.~Vicari,
  JHEP {\bf 0308}, 029 (2003)
  [hep-ph/0307036].

\bibitem{Delamotte:2003dw} 
  B.~Delamotte, D.~Mouhanna and M.~Tissier,
  Phys.\ Rev.\ B {\bf 69}, 134413 (2004)
  [cond-mat/0309101].

\bibitem{Vicari:2007ma}
  E.~Vicari,
  PoS LAT {\bf 2007}, 023 (2007)
  [arXiv:0709.1014 [hep-lat]].

\bibitem{Fukushima:2010ji} 
  K.~Fukushima, K.~Kamikado and B.~Klein,
  Phys.\ Rev.\ D {\bf 83}, 116005 (2011)
  [arXiv:1010.6226 [hep-ph]].

\bibitem{Pelissetto:2013hqa}
  A.~Pelissetto and E.~Vicari,
  arXiv:1309.5446 [hep-lat].

\bibitem{Nakayama:2014sba}
  Y.~Nakayama and T.~Ohtsuki,
  arXiv:1407.6195 [hep-th].

\bibitem{Grahl:2014fna} 
  M.~Grahl,
  arXiv:1410.0985 [hep-th].

\bibitem{Bonati:2014kpa} 
  C.~Bonati, P.~de Forcrand, M.~D'Elia, O.~Philipsen and F.~Sanfilippo,
  arXiv:1408.5086 [hep-lat].

\bibitem{Bazavov:2012qja}
  A.~Bazavov {\it et al.}  [HotQCD Collaboration],
  Phys.\ Rev.\ D {\bf 86}, 094503 (2012)
  [arXiv:1205.3535 [hep-lat]].

\bibitem{Aoki:2012yj}
  S.~Aoki, H.~Fukaya and Y.~Taniguchi,
  Phys.\ Rev.\ D {\bf 86} (2012) 114512
  [arXiv:1209.2061 [hep-lat]].

\bibitem{Cossu:2013uua}
  G.~Cossu, S.~Aoki, H.~Fukaya, S.~Hashimoto, T.~Kaneko, H.~Matsufuru
  and J.~-I.~Noaki,
  Phys.\ Rev.\ D {\bf 87} (2013) 114514
  [arXiv:1304.6145 [hep-lat]].

\bibitem{Buchoff:2013nra}
  M.~I.~Buchoff, M.~Cheng, N.~H.~Christ, H.~-T.~Ding, C.~Jung,
  F.~Karsch, R.~D.~Mawhinney and S.~Mukherjee {\it et al.},
  arXiv:1309.4149 [hep-lat].

\bibitem{Symanzik:1973vg} 
  K.~Symanzik,
  Commun.\ Math.\ Phys.\  {\bf 34}, 7 (1973).

\bibitem{Appelquist:1974tg} 
  T.~Appelquist and J.~Carazzone,
  Phys.\ Rev.\ D {\bf 11}, 2856 (1975).

\bibitem{Aoki:1997er}
 K.~Aoki,
 Phys.\ Lett.\ B {\bf 418}, 125 (1998)
 [hep-ph/9709309].
         
\bibitem{Grahl:2013pba} 
  M.~Grahl and D.~H.~Rischke,
  Phys.\ Rev.\ D {\bf 88}, no. 5, 056014 (2013)
  [arXiv:1307.2184 [hep-th]].

\bibitem{Ejiri:2012rr}
  S.~Ejiri and N.~Yamada,
  Phys.\ Rev.\ Lett.\  {\bf 110}, no. 17, 172001 (2013)
  [arXiv:1212.5899 [hep-lat]].

\bibitem{Ejiri:2014mada}
  S.~Ejiri and N.~Yamada,
  work in progress.

\bibitem{Sato:2013tka} 
  T.~Sato and N.~Yamada,
  PoS LATTICE {\bf 2013}, 430 (2013)
  [arXiv:1311.4621 [hep-lat]].

\bibitem{Brezin:1973jt} 
  E.~Brezin, J.~C.~Le Guillou and J.~Zinn-Justin,
  Phys.\ Rev.\ B {\bf 10}, 892 (1974).

\bibitem{Aoki:2013zfa} 
  S.~Aoki, H.~Fukaya and Y.~Taniguchi,
  PoS LATTICE {\bf 2013}, 139 (2013)
  [arXiv:1312.1417 [hep-lat]].

\bibitem{Collins:1984xc}
 For known exceptions to the decoupling theorem, see, for example,
 J.~C.~Collins,
 Cambridge, Uk: Univ. Pr. ( 1984) 380p

\bibitem{Ukawa:1995tc}
 A.~Ukawa,
 UTHEP-302, C93-06-21.1.

\bibitem{Sato:2014effV}
 A part of calculation of effective potential is shown in
 Tomomi Sato and Norikazu Yamada,
 PoS LATTICE {\bf 2014}, 191 (2014)
 arXiv:1501.06684.

\end{thebibliography}
\end{document}